\documentclass[aps,prd,preprint,superscriptaddress,nofootinbib,showpacs]{revtex4}
\usepackage{slashed}
\usepackage{epsfig,latexsym,cancel,amssymb,amsmath,verbatim,mathrsfs}
\usepackage{color}
\usepackage{graphicx}

\def\ra{\rightarrow}
\def\L{\left(}
\def\R{\right)}
\def\wt{\widetilde}

\def\ld{\lambda}
\def\f{\frac}
\newcommand{\be}{\begin{equation}}
\newcommand{\ee}{\end{equation}}
\newcommand{\bea}{\begin{eqnarray}}
\newcommand{\eea}{\end{eqnarray}}
\newcommand{\ba}{\begin{array}}
\newcommand{\ea}{\end{array}}

\long\def\symbolfootnote[#1]#2{\begingroup%
\def\thefootnote{\fnsymbol{footnote}}\footnote[#1]{#2}\endgroup}

\newcommand{\beq}{\begin{equation}}
\newcommand{\eeq}{\end{equation}}

\begin{document}

\title{Bound States via Higgs Exchanging and Heavy Resonant Di-Higgs }


\author{Zhaofeng Kang}
\email[E-mail: ]{zhaofengkang@gmail.com}
\affiliation{School of Physics, Korea Institute for Advanced Study,
Seoul 130-722, Korea}

\date{\today}

\begin{abstract}

The existence of Higgs boson $h$ predicted by the standard model (SM) was established and hunting for clues to new physics (NP) hidden in $h$ has become the top priority in particle physics. In this paper we explore an intriguing phenomenon that prevails in NP associated with $h$, bound state ($B_h$, referring to the ground state only) of relatively heavy particles $\phi$ out of NP via interchanging $h$. This is well-motivated due to the intrinsic properties of $h$: It has zero spin and light mass, capable of mediating Yukawa interactions; moreover, it may be strongly coupled to $\phi$ in several important contexts, from addressing the naturalness problem by compositeness/supersymmetry (SUSY)/classical scale invariance to understanding neutrino mass origin radiatively and matter asymmetry by electroweak baryogensis. The new resonance $B_h$, being a neutral scalar boson, has important implications to the large hadron collider (LHC) di-Higgs search because it yields a clear resonant di-Higgs signature at the high mass region ($\gtrsim 1$ TeV). In other words, searching for $B_h$ offers a new avenue to probe the hidden sector with a Higgs-portal. For illustration in this paper we concentrate on two  examples, the stop sector in SUSY and an inert Higgs doublet from a radiative neutrino model. In particular, $h$-mediation opens a new and wide window to probe the conventional stoponium and the current date begins to have sensitivity to stoponium around TeV.





\end{abstract}

\maketitle

\section{Force mediator: a new face of Higgs boson}

The main focus of particle physics lies on aspects of the newly discovered member of SM, the Higgs boson $h$. It is commonly believed to be the portal to the mysterious new physics world where the gauge hierarchy problem, dark matter, neutrino mass or/and baryon asymmetry origins may be addressed. Specific to LHC, di-Higgs search is of particular interest since it could help to reveal the Higgs potential~\cite{di Higgs,resonantdih}. 

In this paper we explore the thought-provoking hypothesis that $h$ plays the role of force carrier and mediates new interaction between particles (collectively denoted as $\phi$) out of the NP sector, making them form bound state $B_h$. This hypothesis is well motivated grounded on three basic properties of $h$. First of all, it is a spin-0 particle and thus mediates Yukawa interaction. Next, its mass $m_h\approx 125$ GeV is much lighter than the NP states, which are expected to be around the TeV scale, and thus $h$ can be regarded almost massless. Last but not least, the interacting strengths of $h$ to $\phi$ are unknown but there are convincing examples indicating that they are, or at least can be fairly strong, e.g., in the theories addressing naturalness problem by classical scale invariance and understanding matter asymmetry via electroweak baryogensis, the Higgs field may strongly couple to some new scalar fields so as to trigger classical scale symmetry breaking and strong first-order EWPT, respectively; in particular, in the composite Higgs scenario, $h$, being a pseudo Goldstone boson, is a strong reminiscence~\footnote{For instance, it may bound the exotic spin-1 resonance~\cite{Bellazzini:2012tv} despite not the spin-1/2 top partners, the more robust prediction but with suppressed couplings to $h$ owing to the small composite-fundamental mixing.} of the pion of Hideki Yukawa, which has large couplings with nucleons and thus bound them in nuclei.~\footnote{In the chiral perturbative theory, pions are pseudo Goldstone boson (composite) particles from chiral symmetry breaking. } Therefore, the existence of $B_h$ in NP is in expectation.

\begin{figure}[htb]
\includegraphics[width=3.5in]{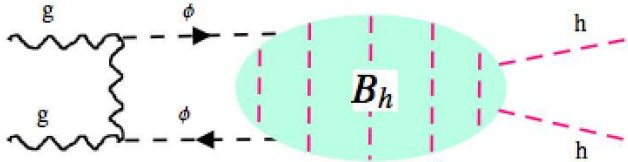}\,\,\,\quad\quad
\caption{Production and decay of bound state $B_h$, which looks like a new resonance dominantly decaying into  di-Higgs boson.}\label{Bhdecay}
\end{figure}

The bound state $B_h$ via Higgs boson exchanging shows a remarkable feature, i.e., it dominantly decays into a pair of force carrier, namely di-Higgs boson. Therefore, as long as the bound state $B_h$ has an abundant production at LHC, we are going to observe a remarkable resonant di-Higgs signature; see Fig.~\ref{Bhdecay}. This new observation is one of the key difference between our paper and the quite old papers which considered Higgs exchange effect restricted to quarkonium, bound state of hypothetical heavy quarks~\cite{Inazawa:1987vk} (or even Higgs-Higgs bound state~\cite{Grinstein:2007iv}). Furthermore, now we already largely pin down the Higgs boson and know it should lead to a new type of interaction other than the gauge interactions, so it is the right time to explore $B_h$ in a wide context of NP. 

\section{General aspects of  $B_h$}

\subsection{Formalism for $B_h$}

 We start with a simplified model which captures the the main features of $B_h$ at LHC. The ingredients include the force carrier $h$ and the constituent $\phi$, which is assumed to be a scalar (complex for the time being) field, along with the Higgs-portal interactions
\begin{align}\label{}
-{\cal L}_h=u_{h\phi\phi} h|\phi|^2+m_\phi|\phi|^2.
\end{align}
The discussions can be easily generalized to other cases, says a fermionic or vector $\phi$. Probably, $\phi$ carries the SM charges such as $SU(3)_C$ and/or $U(1)_Y$, which is important in the production of $B_h$ at LHC.  

In the bound state of $\phi$, the internal motion is nonrelativistic and thus its dynamics can be described by quantum mechanism or concretely, the Schrodinger equation 
\begin{align}\label{yuk}
\L-\f{\triangledown^2}{2\mu_r}+V(\vec r)\R\Psi(\vec r)=E\Psi (\vec r),
\end{align}
where $\mu_r=m_\phi/2$ is the reduced mass and $V$ is the central potential, which, specific to Higgs interchanging, is the Yukawa potential $-\f{\alpha_h}{r}e^{-m_h r}$ with $\alpha_h=u_{h}^2/(16\pi m^2_{ \phi})$~\cite{Petraki:2015hla}. Although the exact analytical solution to Eq.~(\ref{yuk}) is not available, an approximate one can be found based on the scaled Hulthen potential~\cite{improved}
\begin{align}\label{hul}
V_{SH}(r)=-{\alpha_h}\f{R_sm_h\,e^{- R_s m_h r}}{1-e^{- R_s m_h r}},
\end{align}
where $R_s\approx 1.75$~\cite{improved}. Both the standard Hulthen~\cite{Hulthen} and rescaled Hulthen potential resemble the Yukawa potential and admits an exact solution, but the latter is better when the bound state is just marginally formed. Consider the squared bound state wavefunction ($S$-wave) at the origin
\begin{align}\label{spectrum}
 |\Psi_{n}(0)|^2\approx &\f{\epsilon(R_s/D_h)}{n^3}\f{1}{\pi a_0^3}= \f{\L 1-\f{R_s^2}{4D_h^2}\R^{\f{3}{2}}}{n^3}\f{\alpha_h^3 m_B^3}{64\pi},
\end{align}
where $m_B=2m_\phi$ and $a_0\simeq 1/\alpha_h \mu_r$ is the characteristic scale of $B_h$, the Bohr radius; $D_h\equiv m_h^{-1}/a_0 $ is a good measurement of how Coulomb-like the system is. Hereafter we will consider the ground state only, hence dropping the subscript.

The Coulomb limit is $D_h\gg 1$, i.e.,  the screening length $1/m_h$ is much longer than the Bohr radius. If $D_h$ approaches one, the screening effect is strong; the critical condition for the existence of at least one bound state, i.e., the ground state, is $D_h\gtrsim 0.84$~\cite{Rogers,Shepherd:2009sa} (Note that the above approximation may be valid only for $D_h\gtrsim 1$). This condition is fulfilled when 
\begin{align}\label{Dh}
{m_\phi}\gtrsim 0.84\times \f{2}{\alpha_h}m_h\approx 0.7\times\L\f{0.3}{\alpha_h}\R\rm TeV.
\end{align}
Due to the heaviness of force carrier $h$, bound state can exist only for either heavy constitutes or rather strong self-coupling close to the perturbative bound. On the other hand, one can derive an upper bound for the massive coupling by requiring that the lifetime of the bound state should be longer than the time scale of its formation, the inverse of the binding energy~\cite{Hagiwara}, namely
\begin{align}\label{}
\Gamma_{B_h}\lesssim  E_b\Rightarrow \alpha_h\lesssim 1/N_c^{1/3}.
\end{align}
We have used $\Gamma_{B_h}\approx\Gamma_{B_h\ra hh}$ in Eq.~(\ref{Bhh}). $N_c$ is the color factor from $\phi$ and for $N_c=3$ one has $\alpha_h\lesssim 0.7$, while for $N_c=1$ the bound coincides with the perturbativity bound.~\footnote{It may be the right place to make a comment on the possible issue on the unstable force mediator. Naively, in this case the Yukawa potential obtained from one-Higgs boson exchange diagram will be modified as $e^{-m_hr}/(4\pi r)\ra e^{-(m_h+i\Gamma_h/2)r}/(4\pi r)$ with $\Gamma_h$ the Higgs boson width, around 4 MeV in SM. It is much smaller than the Higgs boson mass and thus is of no numerical importance. In other words, as long as the force mediator is sufficiently long-lived compared to the bound state forming time scale, we can treat it as a stable particle.}

\subsection{Resonant di-Higgs signature from $B_h$}

At hadron colliders like LHC, the bound state $B_h$ can be created when the pairly produced $\phi$ have center-of-mass (CM) energy just below the threshold $m_B$. $B_h$ is not stable and overwhelmingly decays into a pair of Higgs boson.~\footnote{In the context of bound state, some authors investigated this but just for exploring the possibility~\cite{Barger}} Therefore, provided a sizable cross section of $B_h$, a clear prediction of new resonance in the di-Higgs channel is furnished. In this subsection we will detail the production and decay of $B_h$.

Let us begin with the annihilation decay of $B_h$ (neglecting the open ``flavor" decay). In general, the partial decay widths of $B_h$ into $XY$ can be calculated in terms of the amplitude of annihilation $ \phi \phi^*\ra XY$ and the bound state wave function at the origin~\cite{Kats:2009bv}
\begin{align}\label{master}
\Gamma_{B\ra XY}=&\f{1}{2m_{B}}\f{N_c}{1+\delta_{XY}}\int d\Pi_2\f{2}{m_{B}} |{\cal M}_{\phi \phi^*\ra XY}|^2 |\Psi(0)|^2,
\end{align}
with $\delta_{XY}$ the statistic factor. For instance, for $A=B=h$ one has 
\begin{align}\label{Bhh}
\Gamma_{B\ra hh}\approx\f{N_c}{16\pi}\f{ |\Psi(0)|^2}{m_B^2} \left[\f{4u_{h\phi\phi}^4}{\L\f{1} {2}m_B^2-m_h^2\R^2}\right]\beta_h,
\end{align}
with $\beta_h=\L 1- 4m_h^2/m_B^2\R^{1/2}$. Since a relatively heavy $\phi$ is required because of Eq.~(\ref{Dh}),  thus $m_\phi^2\gg m_h^2$. Then the squared amplitude (the factor in the squared bracket) can be approximated as $\sim (u_{h\phi\phi}/m_\phi)^4=(16\pi\alpha_h)^2=404\times (\alpha_h/0.4)^2$, a large value. Therefore it tends to dominate over other modes. As a comparison, consider a colored $\phi$, for concreteness in the fundamental representation of $SU(3)_c$ such as stop/sbottom that will be discussed later. Then $B_h$ can decay into $gg$ with width~\cite{Drees:1993uw,Martin:2008sv} 
\begin{align}\label{gg}
\Gamma_{B_h\ra gg}\approx\f{N_c}{16\pi}\f{ |\Psi(0)|^2}{m_B^2} \left[\f{256\pi^2}{9}\alpha_s^2\right]\ll \Gamma_{B_h\ra hh}.
\end{align}
Note that for a scalar $\phi$ with electroweak (EW) charges, the annihilation $\phi\phi^*\ra Z^*/\gamma^* \ra qq$ is $p-$wave suppressed and hence the corresponding $B_h$ production via $q\bar q\ra B_h$ is inaccessible.

On top of those annihilation decay modes via the $u/t$-channels $\phi$ mediation or contact interactions, the decay modes via $s$-channel Higgs mediation may be also important. This is particularly true for the $VV$ modes with $V=W, Z$, because we are considering a TeV scale bound state and thus they obtain the Goldstone enhancement factor $~m_{B}^2/m_V^2\gg1$: ~\footnote{The decay width can be also simply obtained via the $B_h-h$ mixing discussed below.}
\begin{align}\label{VV}
\Gamma_{B_h\ra VV}\approx\delta_V\f{N_c\alpha_h^4}{256\pi}\L\f{m_B}{v}\R^2 m_B,
\end{align}
with $\delta_V=1, 2$ for $V=W$ and $Z$, respectively. Note that there may be additional contributions from other channels if $\phi$ carries $SU(2)_L\times U(1)_Y$ charges, but they are supposed to be subdominant owing to the absence of $u_{h\phi\phi}$ enhancement from $|\phi|^2h$ coupling. As for $\Gamma_{B_h\ra ff}$, dominated by $t\bar t$, is always suppressed for $m_B\gtrsim 0.5$ TeV; the branching ratio typically is $\lesssim 1\%$. In summary, the ratio $\Gamma_{B_h\ra WW+ZZ}/\Gamma_{B_h\ra hh}=3(m_B/v)^2/64\pi$ exceeds 1 for $m_B\gtrsim 2.0$ TeV; for even much heavier $B_h$, its decays become the same as a singlet Higgs boson which decays via its mixing with the SM-like Higgs boson. Whereas for $m_B$ substantially below 2 TeV, it is well justified to set Br$(B_h\ra hh)\simeq100\%$.

In our framework, the Higgs boson and bound state unavoidably mix with each other, and they might have a sizable mixing angle by virtue of the large coupling $u_{h\phi\phi}$. In this case, the bound state (especially formed by the SM singlet or without $S$-wave annihhilation despite of SM charges)  can be produced through gluon-gluon-fusion (GGF)~\cite{Bi:2016gca}. After EW symmetry breaking, $m_{hB}^2$, the mass squared matrix for $h$ and $B_h$ have entries~\cite{Bi:2016gca}
\begin{align}\label{master}
\L m_{hB}^2\R_{11}=&4{\ld} v^2,\quad \L m_{hB}^2\R_{22}\approx m_B^2, \cr
 \L m_{hB}^2\R_{12}=&\L m_{hB}^2\R_{21}=\sqrt{2}\f{|\Psi(0)|}{\sqrt{m_B}}u_{h\phi\phi}\approx \f{\alpha_h^{2}}{2\sqrt{2}} m_B^2,
\end{align}
where $\ld$ is the SM Higgs quartic coupling and $v=174$ GeV. Since $m_B$ is heavy, mixing effect could pull down the SM Higgs boson mass and then we call for a larger $\ld$ than the SM prediction $\ld_{\rm SM}\approx0.13$~\cite{Kang:2012sy}. A conservative bound on $m_B$ and $\alpha_h$ can be derived from requiring the absence of tachyon, which probably signs the condensation of $\phi\phi^*$ and then the pattern of EWSB is modified and therefore the discussions here become invalid.~\footnote{Such a interesting topic has been investigated within the MSSM~\cite{Giudice:1998dj} where the stop bound state is also by exchanging Higgs boson. But that bound state, requiring even much larger $A_t$ coupling, may be not nonrelativisitic, which is different than our object, a nonrelativisitic bound state at LHC.} A more stringent bound is from requiring the mixing angle $\theta_{hB}$ should satisfy $\sin\theta_{hB}\lesssim 0.34$~\cite{mixing}. In our latter studies the mixing angle will be significantly below this bound.

If the mixing angle is very small, $B_h$ can still be produced provided that $\phi$ carries SM charges in particular color. As an example, we identify $B_h$ as the stop bound state, the stoponium. The cross section of $B_h$ production from GGF could be calculated with Eq.~(\ref{gg}) at hand~\cite{Drees:1993uw}:
\begin{align}\label{}
\sigma(gg\ra B)&=f_\zeta(D_h)\f{2}{C_g^2}\f{4\pi^2}{m^3_B}{\cal L}_{gg}(m_B^2)\Gamma_{B\ra gg}\cr
=&f_\zeta(D_h)\f{\pi^2}{96}\f{\alpha_h^3\alpha_s^2}{m_B^2} x^0 \int_{x_0}^1\f{dx}{x}f_g(x)f_g\L\f{x_0}{x}\R,
\end{align}
with $x_0\equiv m_B^2/{ s}$ ($ s$ is the collider energy) and color factor $C_g=8$. For $D_h\gg1$ the factor $f_\zeta(D_h)=\zeta(3)\approx1.2$ comes from the summation over the exciting states $ns$ ($n=1,2,...$). However, for $D_h\sim 1$ only the ground state is accessible and then $f_\zeta(D_h)=1$. However, for a EW charged scalar $\phi$ one cannot expect the $B_h$ production via the Drell-Yan process $q\bar q\ra B_h$ with reasons explained previously.

\subsection{Modifications to the Higgs signatures}


In the simplified model the phenomenology of bound state is closely related to the Higgs signature rates, which in turn restrict the viable parameter space that accommodates $B_h$. If $\phi$ is colored and charged, both the production and radiative decay of $\phi$ will be modified, with amounts\begin{align}\label{}
 \delta r_\gamma &\approx r_{{\rm SM},\gamma}+{\rm sign}(u_{h\phi\phi})\f{d(\phi)Q^2_\phi}{12} \sqrt{2\pi \alpha_h}\f{v}{m_\phi}, \cr
  \delta r_g &\approx r_{{\rm SM},g}+{\rm sign}(u_{h\phi\phi}){C(\phi)}\sqrt{2\pi \alpha_h}\f{v}{m_\phi}, 
\end{align}
with $r_{{\rm SM},g}\approx 0.97$ and $r_{{\rm SM},\gamma}\approx-0.81$. $d(\phi)$ is the dimension of representation of $\phi$ under $SU(3)_c$ and $C(\phi)=1/2$ for $d(\phi)=3$. In a more complete model, it is likely that $\phi$ possesses a partner which also contributes to the above; see examples later. In general, the current data, due to the unprecise knowledge of Higgs couplings while heaviness of $\phi$, has not yielded a stringent constraint yet.


\section{Examples}

\subsection{Large $A_t/\mu$-term $\&$ stoponium/sbottomonium}

 In the UV models $u_{h\phi\phi}$ can be generated in two ways. One is via the usual Higgs portal term and the other one is via a trilinear massive coupling. In NP models there are well motivated examples for both ways, and in the following we present two examples for them one by one.

\begin{figure}[htb]
\includegraphics[width=2.5in]{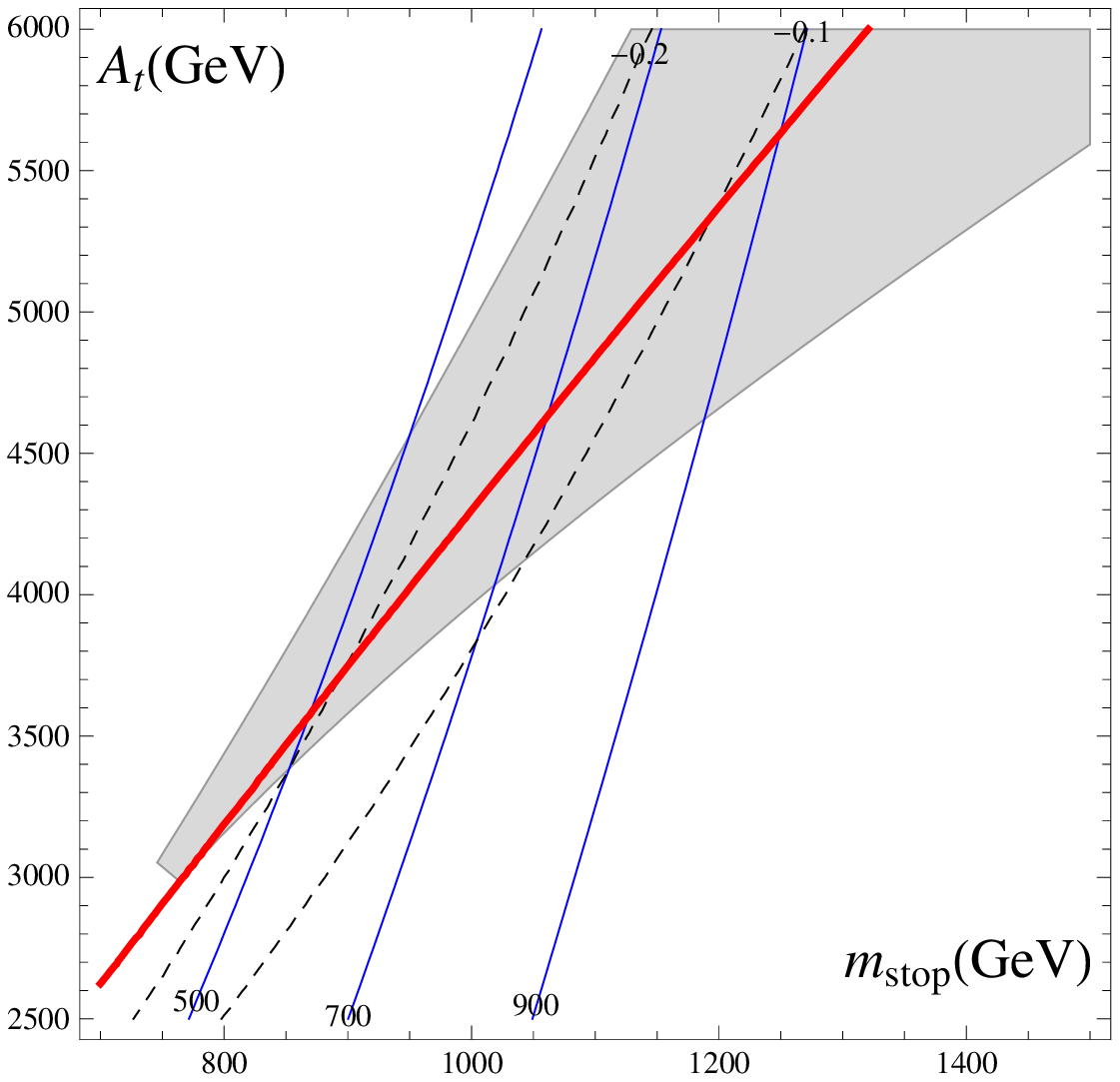}~~~~~~\quad\quad
\includegraphics[width=2.5in]{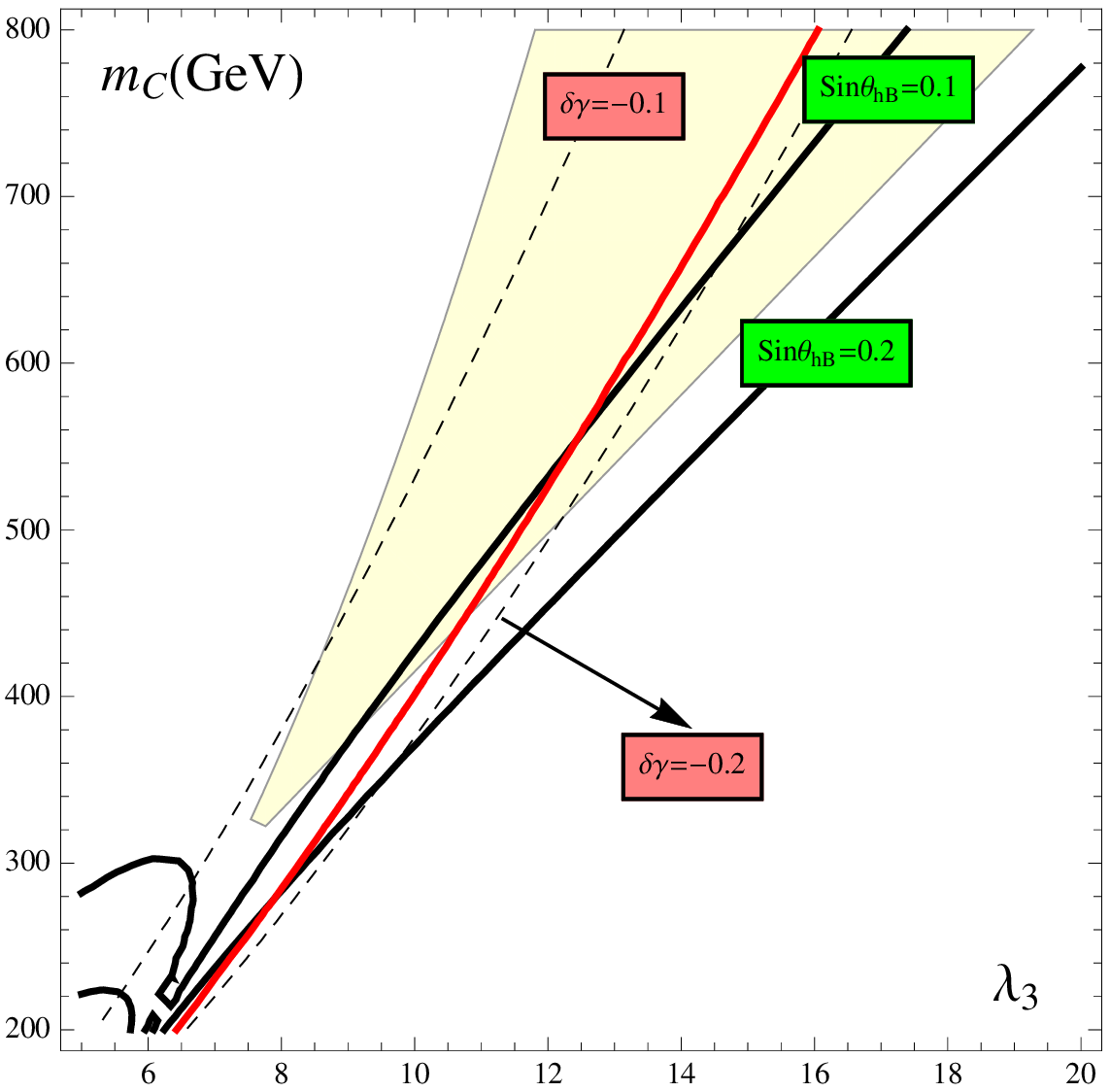}
\caption{Parameter spaces accommodating $B_h$ for two examples. Stop sector (left):  $m_{stop}=m_{LL}=1.3m_{RR}$ and $\ld=0.18$; the shaded region satisfies $m_{\wt t_1}>300$ GeV, $|\sin\theta|<0.2$ and $D_h>0.84$. We show the mass of $m_{\wt t_1}$ (blue lines) and the amount of Higgs diphoton change (dashed lines). IDM (right): the shaded region satisfies $\alpha_h<0.7$ and $D_h>0.84$; $\ld=0.23$. In both plots the red lines, which can be adjusted by taking different $\ld$, label $m_h=125$ GeV.}\label{bound}
\end{figure}

Consider the stop sector in the supersymmetric SMs (SSMs). To mitigate the fine-tuning problem of EW scale caused by the 125 GeV Higgs boson, the stop sector, in particular of the minimal SSM, is strongly favored to have a large trilinear soft SUSY breaking term~\cite{Kang:2012sy} and thus a large coupling $u_{h \wt t_1\wt t_1 }h\wt t_1^*\wt t_1$ is well expected. The stop sector contains three parameters, collected in the stop mass squared
matrix $m^2_{stop}$ (in the basis $(\wt t_R,\,\wt t_L)$):
\begin{align}\label{stop}
m^2_{stop}\approx \left(\begin{array}{cc}
            m_{RR}^2 & m_t  X_t\\
          m_t X_t & m_{LL}^2
                        \end{array}\right),
\end{align}
with $m_{RR/LL}^2$ being free parameters and $X_t=A_t-\mu \cot\beta \approx A_t$ for $\tan\beta=v_u/v_d\gg1$ and a relatively small $\mu$-term for naturalness. $A_t$ characterizes the trilinear soft SUSY breaking term
\begin{align}\label{}
-{\cal L}_{soft}\supset y_t A_t \wt t_L \L v_u+\f{h}{\sqrt{2}}\R\wt t_R^*+h.c.,
\end{align}
where we have assumed an exact decoupling limit of the Higgs bosons. The two mass eigenstates of Eq.~(\ref{stop}) are $\wt t_{1,2}$, which have masses $m_{\wt t_{1,2}}$ and are related to the gauge eigenstates via $\wt t_{L}=\cos{\theta_{t}}\wt t_{1}-\sin\theta_{t}\wt t_2$ and $ \wt t_{R}=\sin{\theta_{t}}\wt t_{1}+\cos\theta_{t}\wt t_2$, with  $\theta_{t}$ the stop mixing angle. With them we derive the massive coupling
\begin{align}\label{}
-{\cal L}_{h\wt t_1\wt t_1}=u_{h \wt t_1^*\wt t_1 } h|\wt t_1|^2~{\rm with} ~u_{h \wt t_1\wt t_1 }\approx \f{m_t A_t}{\sqrt{2}v}\sin2\theta_t. 	
\end{align}
Thus, asides from a large $A_t$, maximal mixing $\theta_t\sim \pi/4$ is required. The typical configuration of stop sector that accommodates $B_h$ is shown in the left panel of Fig.~\ref{bound}; for generality, we do not restrict to the case that stop radiative correction is the only extra source for $m_h$. 


Comments are in orders. First, the sufficiently narrow decay width of $\wt t_1$ can be guaranteed as long as the two-body decay to the lightest sparticle (LSP) is suppressed or even forbidden. For instance, they have almost degenerate mass with the LSP or even they are the NLSP but with a gravitino LSP.  Second, exchanging gluons also contributes to formation of stoponium/sbottomonium, but it is subdominant to the contribution from exchanging Higgs, because typically one has $\alpha_h$ considerably larger than $\f{4}{3}\alpha_s$ under the constraint Eq.~(\ref{Dh}). Third, the Higgs diphoton rate shift, as mentioned before, receive contributions from both stops:
\begin{align}\label{}
\delta r_g\approx \f{1}{4}\L\f{m_t^2}{m_{\wt t_1}^2}+\f{m_t^2}{m_{\wt t_2}^2}-\f{m_t^2A_t^2}{m_{\wt t_1}^2m_{\wt t_2}^2}\R, ~\delta r_\gamma=\f{2}{9}\delta r_g.
\end{align}
Last but not least, sbottomnium is also well motivated from the sbottom sector (see another example in Ref.~\cite{Baek:2015fma}). We  admit a serious little hierarchy problem without a large $A_t$ term to enhance $m_h$. Then, one needs very heavy stops and thus usually, owing to the renomalization group evolution of the Higgs parameter $m_{H_u}^2$, a multi-TeV scale $\mu$-term is necessary to fulfill successful EWSB.  Such a large $\mu$, aided by a large $\tan\beta$ and maximal  sbottom mixing can induce sbottomnium:
\begin{align}\label{}
\mu H_uH_d+y_b Q_3 H_d U^c_3\Rightarrow \f{m_b \tan\beta}{\sqrt{2}v}\mu\sin2\theta_b h|\wt b_1|^2.
\end{align}

\subsection{Can $\phi$ be a dark matter field? }

It is of great interest to consider the situation that $\phi$ is a dark matter field, however, $h$-mediated DM-nucleon spin-independent (SI) scattering excludes this possibility. The cross section can be written as
 \begin{eqnarray}\label{}
\sigma^n_{\rm SI}\approx \f{4\alpha_h}{v^2}\f{m_n^4}{m_h^4} \L \Sigma_{q}f_{T_q}^{(n)}\R^2=3.6\times10^{-6}\times\L\f{\alpha_h}{0.3}\R\rm pb,
\end{eqnarray}
where we have used the values of $f_{T_q}^{(n)}$ presented in Ref.~\cite{Gao:2011ka}. For DM of a few 100 GeVs, the predicted $\sigma^n_{\rm SI}$ exceeds the LUX bound by two orders of magnitude~\cite{LUX}. Despite of the potential cancelation from other contributions in a complete model, we are interested in a more viable and natural scenario: The dark sector contains some heavier states other than the DM candidate; some of them are long-lived due to the narrow decay width into the lighter dark states, thus being the candidate of $\phi$.

A good case in point is the inert Higgs doublet $\Phi_1$ ($\Phi_2$ is the SM Higgs doublet) from the celebrated radiative neutrino model of Ma~\cite{E. Ma}, where the singlet Majorana fermion $N$ (single family for our purpose) is identified as the DM candidate~\cite{fermionicDM} and $\Phi_1$ provides the candidate for $\phi$. The relevant terms are collected in the following
 \begin{align}\label{}
-{\cal L}_{\rm Ma}&=\ld_3|\Phi_1|^2|\Phi_2|^2+\ld_4|\Phi_1^\dagger\Phi_2|^2\cr
+&\f{\ld_5}{2}\left[(\Phi_1^\dagger\Phi_2) ^2+h.c.\right]
+\L y_N\bar l \Phi_1 P_R N+c.c.\R,
\end{align}
where $y_N$ is small to make $\phi$ slowly decay. The mass spectrum of $\Phi_1=(C^+,\, (S+iA)/\sqrt{2})^T$ is
 \begin{align}\label{}
m_A^2=m_S^2-2\ld_5 v^2,\quad m_C^2=m_S^2-(\ld_4+\ld_5)v^2,
\end{align}
with $m_S$, the mass of $S$, a free parameter. The trilinear couplings involving a single $h$ can be written as
\begin{eqnarray}\label{}
-{\cal L}_h&\supset \sqrt{2}\ld_3 v\, hC^+C^-+\f{v}{\sqrt{2}}\L\ld_3-\f{m_C^2-m_S^2}{v^2}\R hS^2
\cr
&+\f{v}{\sqrt{2}}\L\ld_3-\f{m_C^2-m_A^2}{v^2}\R hA^2.
\end{eqnarray}
We choose $C=\phi$. Let us explain why $S/A$ cannot be the DM candidate. To ensure $h$ very weakly coupled to DM$^2$ but strongly to $C^+C^-$, a large mass splitting between DM and $C$ is necessary; in turn, the decay $C\ra {\rm DM}+W$ is rendered too fast, thus $C$ failing to be $\phi$.

In the absence of DM constraints, $\ld_{3,4,5}$ can be large, only loosely constrained by perturbativity, $\lesssim8\pi$~\cite{Arhrib:2013ela}. Consider $|\ld_4|,\,|\ld_5|,1\,\ll \ld_3$ ($\ld_3>0$ for vacuum stability), which gives rise to a degenerate spectrum and thus naturally passes the EW precision test. Now we have $\alpha_h=\L\ld_3 v/m_S\R^2/8\pi$. Combining with the bound Eq.~(\ref{Dh}), a large $\ld_3$ is required: $\ld_3=8.5\times \L \f{D_h}{0.84}\R\L\f{0.5}{\alpha_h}\R^{\f{1}{2}}$. It is well motivated to trigger a strong first-order electroweak phase transition (SFOPT). Optimistically, the strength of SFOPT is estimated to be~\cite{Chowdhury:2011ga,Chung:2012vg}
\begin{align}\label{}
v(T_c)/T_c\sim 4(\ld_3/2)^{3/2}/6\pi \ld_{\rm SM}\sim 10,
\end{align}
which is strong enough for successful EW baryogensis. With that large $\ld_3$, $C$ decreases the Higgs to diphoton rate by an amount about $\f{2}{r_{\rm SM,\gamma}}\f{\pi}{3} \f{\alpha_h}{\ld_3}=10\%\times\f{D_h}{0.84}\sqrt{\f{\alpha_h}{0.5}}$. Several aspects of this bound state are demonstrated in the right panel of Fig.~\ref{bound}.

Actually, the key point of the above model is nothing but the usual Higgs portal $\eta|\Phi_2|^2|\phi|^2$ with $\eta\gg1 $. Such a strongly interacting portal is crucial in any models  (not only in the IDM above) for SFOPT via bosonic thermal loops, and so does in the models for triggering classical scale symmetry breaking via new bosonic degrees of freedom~\cite{Guo:2014bha}. In other words, in such a large kind of models which are well motivated by NP one can expect $B_h$.

\subsection{Current status of $B_h$ at LHC}

\begin{figure}[htb]
\includegraphics[width=4.0in]{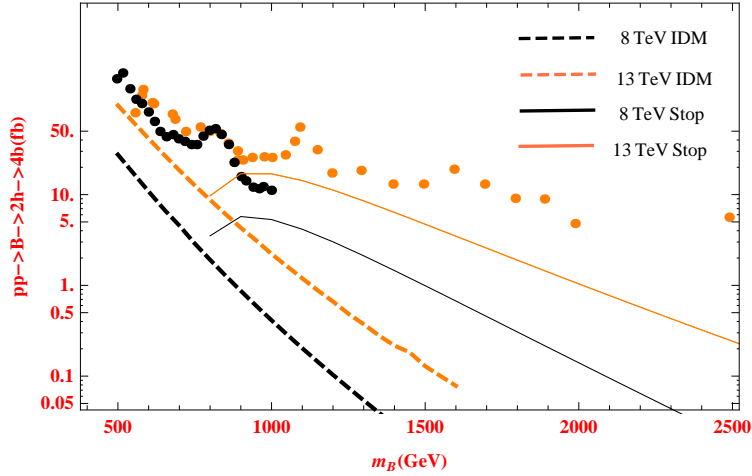}\,\,\,\quad\quad
\caption{Current status of resonant di-Higgs: $B_h$ from stop (solid lines) and IDM (dashed lines). The orange and black dots are the LHC 13 TeV and 8 TeV data, respectively.}\label{diH}
\end{figure}

Searching for $B_h$ (via di-Higgs resonance) offers a new avenue to probe the hidden sector with a Higgs-portal. Recently, resonant di-Higgs signature $pp\ra X\ra hh$ has been extensively searched at 8 $\&$ 13 TeV LHC in various channels, $b\bar b \gamma\gamma$~\cite{Aad:2015xja,Khachatryan:2016sey,CMS:2016guv}, $b\bar b\tau\bar \tau$~\cite{CMS:2016guv} and $4b$, resolved or boosted~\cite{4b,Khachatryan:2016cfa}. The best sensitivity has reached ${\cal O}(10)$ fb near the TeV region for $X$  (in 4$b$ channel). To end up this paper, we demonstrate the present sensitivity to $B_h$ at LHC, with those from the stop sector and IDM as examples; see Fig.~\ref{diH}. It is seen that for the stop case (solid lines) with $\alpha_h=0.7$, the current LHC 13 TeV date begins to have sensitivity to $B_h$ around 1.0 TeV; while for the IDM case (dashed lines), the production of $B_h$ is via Higgs-$B_h$ mixing (taking a relatively large value $\sin\theta_{hB}= 0.15$), we still have to wait for more data to yield a bound. Anyhow, a good sensitivity to $B_h$ probably would be delayed to the 14 TeV LHC. And if a heavy di-Higgs resonance is observed then, $B_h$ would be a very competitive candidate. 

\section{Conclusions}

 Bound state via the SM Higgs boson exchanging is well expected in NP, for instance in supersymmetry and radiative neutrino models. This kind of bound state is bound to show up in the resonant di-Higgs channel and maybe in the Higgs precision tests.

\section{Acknowledgements}

ZK is in debt to Jinmian Li, who offered great help in calculating the luminlocity.

\appendix


\end{document}